\documentclass[aps,twocolumn,preprintnumbers,prl,amsmath,
amssymb,amsfonts,superscriptaddress,floatfix]{revtex4}
\usepackage{amsfonts}
\usepackage{amssymb}
\usepackage{amsmath}
\usepackage{ulem}
\usepackage{color}
\usepackage{latexsym}
\usepackage{graphicx}
\usepackage{graphics}
\usepackage{floatflt}
\usepackage{epsfig}
\usepackage{overpic}
\usepackage{soul} 
\usepackage[latin1]{inputenc}

\newcommand{\red}{\textcolor{black}}

\newcommand{\blue}{\textcolor{black}}
\newcommand{\be}{\begin{equation}}
\newcommand{\ee}{\end{equation}}
\newcommand{\bea}{\begin{eqnarray}}
\newcommand{\eea}{\end{eqnarray}}

\newcommand{\beginsupplement}{%
        \setcounter{table}{0}
        \renewcommand{\thetable}{S\arabic{table}}%
        \setcounter{figure}{0}
        \renewcommand{\thefigure}{S\arabic{figure}}%
     }
\def\nn{\nonumber}
\def\lb{\label}

\def\G{\Gamma}

\newcount\bozza \bozza=0
\ifnum\bozza=1
\newdimen\shift \shift=-2truecm
\def\lb#1{%
{\label{#1}\rlap{\kern\shift{$\scriptstyle#1$}}}}
\else\def\lb#1{\label{#1}} \fi

\begin{document}
\title{Nematic spectral signatures of the Hund's metal}
\author{Laura Fanfarillo}
\affiliation{Scuola Internazionale Superiore di Studi Avanzati (SISSA), Via Bonomea 265, 34136 Trieste, Italy}
\affiliation{Department of Physics, University of Florida, Gainesville, Florida, USA}
\author{Angelo Valli}
\affiliation{Institute for Theoretical Physics, Vienna University of Technology, 1040 Vienna, Austria}
\author{Massimo Capone}
\affiliation{Scuola Internazionale Superiore di Studi Avanzati (SISSA), Via Bonomea 265, 34136 Trieste, Italy}

\begin{abstract}
We show, by means of dynamical mean-field theory calculations, that the experimental fingerprints of the nematic order in iron-based superconductors are intrinsically connected with the electronic correlations in the Hund's correlated metallic state and they can not be accessed via a renormalized quasiparticle picture. In particular, our results show that: 
(i) in a metal in which correlations are dominated by the Hund's coupling the nematic ordering does not produce a rigid energy shift in the photoemission spectra, but a much richer spectral weight redistribution which mirrors the experimental results; (ii) the nematic ordering is characterized by an orbital-selective coherence induced by the Hund's physics in agreement with the experimental picture. 
\end{abstract}

{\bf \maketitle }

Several quantum materials display a large anisotropy in the electronic properties which has been identified as a signature of electronic nematic order where the in-plane rotational symmetry of the electron fluid is broken making $x$ and $y$ directions inequivalent. 
This appears to be an ubiquitous phenomenon in iron-based superconductors (FeSC) \cite{Fernandes_NatPhys2014, Bohmer_review2016, Gallais_review2016,Fernandes_Nature2022}.

Among the different experimental probes, a crucial piece of information can be obtained by Angle-Resolved Photoemission Spectroscopy (ARPES)  exploiting different polarization of light to selectively probe different iron orbitals.  This is particularly relevant in light of the prominent role of the orbital degree of freedom in the electronic structure of FeSC \cite{Yi_npjQM2017}. ARPES studies reveal that the band dispersion in the nematic phase is characterized by a momentum-modulated energy splitting of the $xz$ and $yz$ orbitals. Early studies mainly focused on  FeSe \cite{Suzuki_PRB2015, Fanfarillo_PRB2016, Zhang_PRB2016, Yi_PRX2019} where nematicity emerges in the absence of long-range magnetic order, but the same pattern has been \red{later} observed in BaFe$_2$As$_2$ \cite{Pfau_PRL2019}. 
\red{Only recently in-depth investigations revealed that the nematic order does not only affect the band dispersion, but also the incoherent spectral weight redistribution \cite{Cai_CPB2020, Pfau_PRB2021_122, Pfau_PRB2021_FeSe}. Interestingly, the $xz/yz$ orbital differentiation appears at high-frequencies with an opposite sign with respect to the orbital splitting of the bands at the Fermi energy.}

A large body of experimental evidence suggests that the origin of the nematic phase is an electronic instability inducing anisotropy in the B$_{2g}$ channel \cite{Gallais_review2016, Bohmer_review2016}. Both Ising spin-nematic models and orbital-fluctuation based approaches have been proposed and extensively discussed in the literature \cite{Fernandes_NatPhys2014, Fernandes_Nature2022}. Regardless of the origin of the nematic instability, the characterization of the nematic phase of FeSC as emerging from experiments, clearly calls for a theoretical scheme which includes the sizeable electron-electron interactions and the consequent correlation effects, which are responsible of non-trivial redistribution of spectral weight at different energy scales, as well as the presence of orbital-selective coherence in the many-body nematic state. The crucial role of electron-electron interactions does not come as a surprise after several investigations demonstrating that peculiar, orbital-selective, correlation effects dominate the normal-state of the FeSC \cite{DeMedici_PRL2014}.

The identification of orbital-selective Mott physics is one of the outcomes of a theoretical path which has  clarified the central role of the Hund's coupling in the multiorbital systems \cite{Yin_NatMat2011, DeMedici_PRB2011, Werner_NatPhys2012, Hardy_PRL2013, DeMedici_PRL2014, Maletz_PRB2014, Yi_NatComm2015, McNally_PRB2015, Backes_PRB2015, Hardy_PRB2016, Lafuerza_PRB2017, Watson_PRB2017}.
In this framework, the normal phase has been identified as a Hund's metal, a strongly correlated bad metallic state with distinctive correlation properties \cite{Georges_Review2013, Werner_PRL2008, Haule_NJP2009, DeMedici_PRL2009, Ishida_PRB2010, Liebsch_PRB2010, Yin_NatMat2011, DeMedici_PRL2011, Yu_PRB2012, Bascones_PRB2012, Lanata_PRB2013, DeMedici_PRL2014, Fanfarillo_PRB2015, DeMedici_Chapter2015, Stadler_AP2018, Isidori_PRL2019, Mezio_PRB2019,Richaud2021} that interpolate between a description of incoherent and localized atomic states at high energy and one of coherent states at low energy \cite{Fernandes_Nature2022, Lafuerza_PRB2017,Tong_NatMat2019, Xingye_NatPhys2022, Kreisel_Frontiers2022, Fanfarillo_NatPhys2022} 
and in which orbital-selective physics emerges as a consequence of an effective decoupling between orbitals in a high-spin state \cite{DeMedici_PRL2014, Fanfarillo_PRB2015, Capone_NatMat2018, Mezio_PRB2019}. 

While we have a fairly good understanding of the role of electronic correlations in the normal state, much less in known about broken-symmetry phases. The link with the nematic phase has been touched upon in \cite{Fanfarillo_PRB2017, Yu_PRL2018} using slave-spin mean-fied theories which describe the low-energy excitations as Fermi-liquid quasiparticles. The analysis of the nematic susceptibility in the correlated regime reveals that, if the symmetry between $xz$ and $yz$ orbitals is explicitly broken, the nematic order is strongly affected by Hund's driven correlations that stabilize configurations with small occupation imbalance between the $xz$/$yz$ orbitals \cite{Fanfarillo_PRB2017} like e.g. the sign-change nematic order experimentally observed in FeSC \cite{Suzuki_PRB2015, Fanfarillo_PRB2016, Zhang_PRB2016, Yi_PRX2019,Pfau_PRL2019}.

While these slave-particle studies \cite{Fanfarillo_PRB2017, Yu_PRL2018} provide a reliable description of the nematic reconstruction of the band dispersion, they can not access other fundamental properties of the electronic state, such as the spectral weight transfer, which can involve different energy scales, and the coherence of the electronic states with different orbital character. \red{Moreover, a trivial extension of the low-energy results provided by those studies to higher energies would produce a picture completely opposite to what recently found by ARPES experiments that revealed an opposite trend in the orbital differentation observed close to the Fermi energy with respect to what realized at higher frequencies.}

\red{This calls for an analysis of the interplay between electronic correlations and nematic order beyond the Fermi-liquid picture exploited by slave-particles studies. 
In this work we address this issue by analyzing the nematic spectral signatures of the Hund's metal using a theoretical description that contains the dynamical correlations of the Hund's metal as treated within Dynamical Mean-Field Theory (DMFT)\cite{DMFT_review}. }

Our main result is that the nematic spectral weight transfer in the Hund's metal \red{presents specific feature that differentiate its phenomenology from the nematicity realized in} an ordinary correlated metal characterized by similar effective mass and density of states at the Fermi level. 
In an ordinary correlated metal the nematic order produces a rigid symmetric shift of the $xz/yz$ orbitals spectral weight around the Fermi energy. On the contrary, the Hund's metal experiences strong orbital-selective frequency modulation of the spectra as a result of a nematic symmetry breaking.
\red{The frequency dependence of the orbital anisotropy appears non monotonic and controlled by multiple energy scale, in contrast to what happen in an ordinary correlated metal in which the characteristic energy of the nematic order dynamic is controlled uniquely by the screened Coulomb repulsion $U$.} \blue{A clear differentiation of the correlations effects encoded in the self-energy renormalization at low frequencies and high frequencies is at the origin of the distinctive features of the nematic spectra of the Hund's metal.}

Our work identifies clear signatures of Hund's metal nematicity that explain the main features of the non-trivial ARPES spectra recently observed in \cite{Pfau_PRB2021_122, Pfau_PRB2021_FeSe}, thereby proving that the comprehensive experimental picture that emerges combining the observations of a reconstruction of band dispersion and the spectral weight transfer on the nematic phase of FeSC can be fully understand only accounting for the interplay of nematicity and Hund's metal physics.\\

In order to study the effect of electronic correlations at a reasonable computational cost we consider a minimal model, already used in \cite{Fanfarillo_PRL2020}, which accounts for the main features of the electronic structure of FeSC and for the electron-electron correlations induced by the combined effect of the Hubbard repulsion, $U$, and the Hund's coupling $J_H$. The kinetic Hamiltonian is given by a three-orbital tight-binding model adapted from \cite{Daghofer_PRB2010}, $ H_{0} = \sum_{{\bf k}\sigma} \sum_{\mu \nu }T^{\mu\nu}({\bf k}) c^{\dagger}_{{\bf k}\mu\sigma} c^{\phantom{\dagger}}_{{\bf k}\nu\sigma} $ where $\mu,\nu$ are orbital indices for the $yz$, $xz$, $xy$ orbitals. $c^{\dagger}_{{\bf k}\mu\sigma}$ ($c^{\phantom{\dagger}}_{{\bf k}\nu\sigma}$) is the fermionic operators that creates (annihilates) an electron in orbital $\mu$, with momentum ${\bf k}$ and spin $\sigma$. The set of parameters chosen results in a bare bandwidth $W=1.6$ eV and reproduce qualitatively the shape and the orbital content of the Fermi surfaces typical of the FeSC family, namely two hole-like pockets composed by $yz$-$xz$ orbitals at the $\G$ point and two elliptical electron-like pockets formed by $xy$ and $yz/xz$ orbitals centered at the $X/Y$ point of the 1Fe-Brillouin Zone \cite{suppl}.
Local electronic interactions are included considering the multiorbital Kanamori Hamiltonian which parametrizes the electron-electron interactions in term of a Hubbard-like repulsion $U$ and an exchange coupling $J_H$ favoring high-spin states \cite{Georges_Review2013}. 

We account phenomenologically for the nematic order by adding to the Hamiltonian a bare nematic perturbation $\sim \eta (n_{yz} - n_{xz})$ , $\eta>0$. Rather than looking for a spontaneous nematic symmetry breaking in our simplified model, or considering a specific low-energy origin for the same instability, we focus on the role of the electronic correlations and in particular on their effect on spectral properties in the nematic phase.  

We compute the nematic orbital spectral functions using the full orbital and frequency-dependent DMFT self-energy $\Sigma_{\mu\mu}(i \omega_n)$, where $\omega_n$ is the $n$-th fermionic Matsubara frequency. \red{The orbital dependence of the self-energy leads to a self-consistent renormalization of the nematic splitting. The effect of correlations within a Fermi-liquid quasiparticle picture, encoded in the quasiparticle weight $Z_{\mu}$, can be exctracted from the DMFT self-energy behavior at low-frequency as $Z_{\mu}= (1 - \partial \Im \Sigma_{\mu\mu}/\partial \omega_n)^{-1}$.}  We use an exact diagonalization solver at zero temperature \cite{Capone_PRB2007, Weber_PRB2012, Amaricci_CPC2022, suppl} at a density of four electrons in three orbitals per site, that reproduces the low-energy electronic structure with hole and electron pockets including the momentum-dependence of the nematic splitting \cite{Fanfarillo_PRB2017}. 

\begin{figure*}
\includegraphics[width=0.98\textwidth]{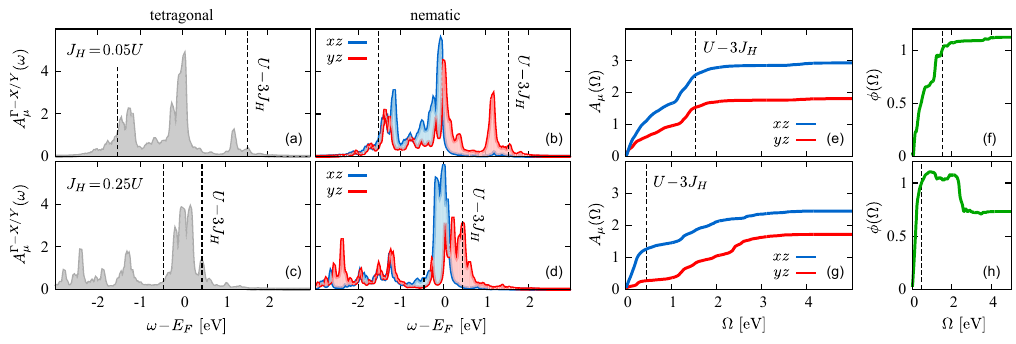}
\caption{Spectral function analysis. Orbital spectral function computed using orbital and frequency-dependent DMFT self-energy for $U \sim W$ in an ordinary correlated metal, $J_H=0.05U$ (a,b) and in a Hund's metal, $J_H=0.25U$ (c,d). In (a,c) we report the spectra computed in the tetragonal phase, using $\delta\epsilon =0$ and in (b,d) the ones for the nematic phase, using $\delta\epsilon=50$~meV.  The nematic order introduces $xz/yz$ orbital differentiation both in the low- and high-$J_H$ regimes, however in the ordinary correlated metal (b) this appears as rigid shift of the $xz/yz$ orbital slightly below/above the Fermi energy, while in the Hund's metal (d) we find an orbital-selective modulation in frequency with the $xz$ orbital remaining closer to $E_F$ and the $yz$ weight moving at higher energy. (e,g) Frequency integration of the orbital nematic spectra $A_{\mu}$ shown in (b,d) as a function of the cut-off $\Omega$. (f,h) Frequency dependence of the nematic parameter $\phi = A_{xz} - A_{yz}$. $\phi$ grows monotonically up to energy $\Omega \sim U$ in the low-$J_H$ regime. In the Hund's metal instead it grows rapidly at low frequency reaching its maximum at $\Omega \sim U-J_H$, it is then suppressed and saturates to a constant value at high frequency. The dashed vertical lines in each panel denote the energy scale $U-3J_H$.}
\label{fig:spectra_nematic}
\vspace{-0.3cm}
\end{figure*}

One of our main goals is to assess the effects of dynamical correlations induced by the Hund's coupling on the spectral properties in the nematic phase. In order to highlight these effects, we focus on two correlated regimes having similar values of $Z_{\mu} \sim 0.3$ in the tetragonal phase (see Table~\ref{table:Z}), but characterized by different values of the Hund's coupling: $J_H =0.05 U$, $U \sim W$ define a ordinary correlated metal, while for $J_H = 0.25 U$, $U \sim W$ we are inside the Hund's metal regime. The spectra in the tetragonal state for both cases are shown in Fig.~\ref{fig:spectra_nematic}(a,e). \red{In order to directly compare the outcome of our calculations with ARPES experiments \cite{Pfau_PRB2021_122, Pfau_PRB2021_FeSe}, we plot the spectral function $A_{\mu}({\bf k}, \omega)$ integrated between the high-symmetry points $\Gamma$-X/Y of the 1Fe-Brillouin Zone, for the $yz/xz$ orbital.}

\begin{table}[tbh]
\begin{center}
\begin{tabular}{c c c c c }
             & \quad \ & Tetragonal & \quad \quad \ & Nematic  \\ 
\hline 
\hline
$J_H=0.05 U$ &     & $Z_{xz/yz}=0.38$ &    & $Z_{xz}=0.45$, $Z_{yz}=0.18$\\   
$J_H=0.25 U$ &     &$Z_{xz/yz}=0.25$ &    & $Z_{xz}=0.35$, $Z_{yz}=0.18$\\
\hline
\end{tabular}
\caption{Quasiparticle renormalization factors extracted by orbital-dependent DMFT self-energy. Starting from degenerate values of $Z_{xz/yz}$ in the tetragonal phase, orbital differentiation developes in the nematic state with the $xz$ orbital remaining more coherent than the $yz$. }
\label{table:Z}
\end{center}
\vspace{-0.3cm}
\end{table}

For small $J_H/U$ we recover the familiar Mott-like behavior where Hubbard bands develop on an energy scale which approaches $U$ in the strong-coupling limit. For larger $J_H/U$ we find that the spectral weight reshuffling involves also a significantly smaller energy scale  $\simeq U-3J_H$, which emerges as the effective charge-charge repulsion in the Kanamori model \cite{Isidori_PRL2019}. \\

In what follows we anayze the nematic spectra and show that the Hund's metal state is affected by the nematic ordering in a much more subtle way with respect to the low-$J_H/U$ regime. To some extent, the main difference with respect to a ordinary correlated metal is that the low-energy scale where the quasiparticles live is not decoupled from the high-energy ($\sim U$) features that evolve into the Hubbard bands.  In the Hund's metal the spectral weight redistribution due to local interactions accumulates also in a narrower energy window around the Fermi energy \cite{DeMedici_PRB2011, Werner_NatPhys2012, Backes_PRB2015, Stadler_AP2018, Fanfarillo_PRL2020}. 
This feature emerged already as crucial to boost boson-mediated superconductivity in Hund's metal \cite{Fanfarillo_PRL2020} and it is expected to critically affect the interplay between local electronic interactions and other low-energy instabilities including the nematic order.

In Fig.~\ref{fig:spectra_nematic}(b,d) we show the nematic orbital spectra for our two choices of parameters: $J_H = 0.05 U$ (ordinary correlated metal) and $J_H = 0.25 U$ (Hund's metal). In both regimes, the nematic order does not alter the overall energy window in which the spectral weight is distributed with respect to the tetragonal phase, however it produces a differentiation in the $xz/yz$ orbital coherence.
\red{Interestingly, while the differentiation produced in the quasiparticle renormalization factors $Z_{\mu}$, listed in Table~\ref{table:Z}, is similar in the two correlated regimes, the orbital-dependent spectral weight redistribution appears quite different.}
By comparing the spectra of the ordinary correlated metal in the tetragonal and nematic phase, panels (a,b),  we find the $xz/yz$ orbitals weights rigidly and symmetrically shifted around the Fermi energy $E_F$. In the Hund's metal instead, panels (c,d), the orbital weight redistribution is not symmetric and thus the orbital anisotropy of the spectra is much more pronounced. The $xz$ spectral weight remains closer to $E_F$ than the $yz$, see larger $xz$ peak close to $E_F$ in panel (d). The $yz$ orbital weight is transferred, instead, to much higher energies where the Hubbard bands are located.
It is worth noticing that this result is due to the dynamic properties of the Hund's metal and cannot be inferred by the analysis of the quasiparticle renormalization factors $Z_{xz/yz}$. Those can only accounts for a rigid shift of the $xz/yz$ orbital spectral weight around $E_F$, while they cannot reproduce the orbital selective behavior at higher frequencies observed in the Hund's metal regime.

To better visualize the frequency dependence of the orbital anisotropy of the spectral weight shown in Fig.~\ref{fig:spectra_nematic}(b,d), we perform a frequency integration over the occupied states of the spectral function on a window of amplitude $\Omega$
\begin{equation}
A_{yz/xz}^{\Gamma-X/Y} (\Omega) = \int_{-\Omega}^{0} d \omega \int_{\Gamma}^{X/Y} d{\bf k} \ A_{yz/xz}({\bf k}, \omega) %f(\omega)    
\end{equation}
and analyze the dynamics of the nematic spectra parameter $\phi(\Omega)= A_{xz} (\Omega)-A_{yz} (\Omega)$.
In Fig.~\ref{fig:spectra_nematic}(e,f) and (g,h) we show respectively $A_{xz/yz} (\Omega)$ and $\phi(\Omega)$ for the low- and high-$J_H/U$ regimes.
In the ordinary correlated metal the anisotropy of the $A_{xz/yz}$ spectral functions, panel (e), and, as a consequence, the nematic parameter $\phi$, panel (f), grow monotonically at an essentially constant rate as we increase the integration window $\Omega$ until it reaches values of order $U$. On the other hand, in the Hund's metal the orbital spectral function $A_{xz/yz}$, panel (g), rapidly deviate from each other for small values of $\Omega$, but they even get closer at higher energy due to the orbital frequency modulation visible in panel (d). As a consequence, the energy dependence of  $\phi$, shown in panel (h), is characterized by a fast growth at low energy, while at frequency $\Omega > U-3J_H$ the nematic order decreases as a function of $\Omega$ and saturates at higher frequencies to a value approximately $\sim 0.75$ of the maximum.

\begin{figure}
\includegraphics[width=0.47\textwidth]{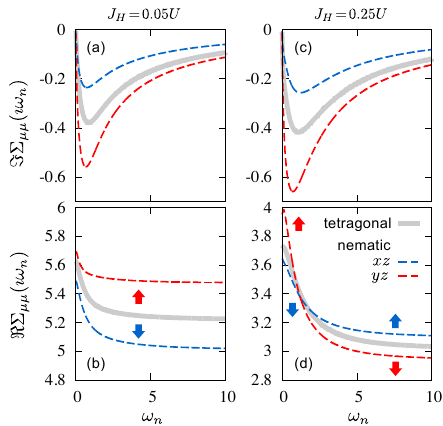}
\caption{Matsubara orbital-resolved self-energy in the tetragonal (solid grey lines) and nematic phase (color dashed lines) for $J_H=0.05U$ (a,b) and $J_H=0.25U$ (c,d). 
The slope of the imaginary yields the anisotropy of the orbital coherence $Z_{xz}>Z_{yz}$ 
and show the same qualitative behavior regardless the value of $J_H$.  
Instead $\Re\Delta\Sigma_{\mu\mu}$, obtained as the difference between the nematic and tetragonal self-energies for each orbital, is strongly affected by the Hund's coupling. 
While having always opposite sign for the $xz$ and $yz$ orbitals, 
at $J_H=0.05U$ has the same sign as the nematic perturbation and is weakly frequency dependent (i.e., it results in a nearly rigid shift), 
while at $J_H=0.25U$ it changes sign between low- and high-energies, reflecting the complex spectral weight redistribution in the Hund's metal. }
\label{fig:self_nematic}
\vspace{-0.3cm}
\end{figure}

\red{To clarify the physical origin of the behavior of the nematic spectral weight in the Hund's metal we look at the self-energy on the imagainary Matsubara axis in the two different correlated case, this allows us to disentangle the interaction-driven contribution to the single particle spectra shown in Fig.~\ref{fig:spectra_nematic}(b,d). In Fig.~\ref{fig:self_nematic} we  show the imaginary and real parts of the self-energy in the two correlated regimes and contrast the nematic results with the tetragonal ones. The imaginary part, panels (a,c), whose low-frequency behavior provides estimate of the quasiparticle weight $Z_{\mu}$, has the same qualitative behavior in the two cases, showing a finite value of $Z_{\mu}$ which becomes larger for the $xz$ orbital and smaller for the $yz$ one, see Table~\ref{table:Z} . This effect recovers what found within Fermi-liquid quasiparticle approximation by previous investigations via slave-particle methods \cite{Fanfarillo_PRB2017, Yu_PRL2018}. 
The real part of the self-energy shows instead a remarkable difference within the two correlated cases. While for small $J_H/U$, panel (b), the orbital-resolved change of the self-energy follows the nematic perturbation at every frequency, for sizable $J_H/U$, panel (d), there is a change of sign at some finite frequency, meaning that only the low-energy part of the spectrum follows the sign of the nematic perturbation, while the high-energy spectrum moves in the opposite direction. This is a clearly nontrivial result that cannot be deduced from the analysis of the orbital anisotropy of the quasiparticle weight $Z_{\mu}$, and required instead a proper inclusion of correlations at DMFT level in order to differentiate low- and high-frequency behaviors. 
The frequency modulation characterized the real part of the self-energy in the Hund's metal regime is at the origin of the remarkably different redistribution of the orbital spectral weight in the low- {\it vs} high-$J_H/U$ regimes shown in Fig.~\ref{fig:spectra_nematic}(b,d).}

By comparing our theoretical findings to the recent ARPES results on the orbital coherence in the nematic phase of FeSC \cite{Pfau_PRB2021_122, Pfau_PRB2021_FeSe} we argue that the experimental spectra show clear signatures of Hund's metal nematicity characterized by an orbital-selective spectral weight redistribution qualitatively compatible with Fig.~\ref{fig:spectra_nematic}(d) and a frequency-modulated nematic order with an intermediate-energy contribution which partially cancels the low-energy signal as shown in Fig.~\ref{fig:spectra_nematic}(h). \red{The peculiar behavior of the nematic spectra is traced down to a frequency-modulated orbital differentiation of the real part of the self-energy in the Hund's metal that displays opposite sign at low {\it vs} high frequencies as shown in Fig.~\ref{fig:self_nematic}(d).}\\

In conclusion we have analyzed the effects of electronic correlations including the Hund's exchange coupling on the nematic phase of a multiorbital model for iron-based superconductors. Comparing results for small values of the Hund's coupling, which behave as a standard Mott-Hubbard system, with large values of $J_H$, that drive the system into a Hund's metal, we are able to demonstrate that the effects of strong correlations on the nematic order can not be describes merely in terms of an orbital-dependent quasiparticle weight reflecting the differentiation between the  $xz$ and $yz$ orbitals. Rather, the full frequency dependence of the interaction effects must be taken into account. 

Our analysis allows us to clearly identify the distinctive signatures of the nematic spectra of a Hund's metal characterized by a frequency dependence of the nematic order \red{originated by an opposite orbital differentiation of the self-energy at low and high frequencies.}

Our results are in excellent agreement with experimental ARPES spectra that show how the nematic reconstruction of band dispersion is accompanied by a non-trivial spectral-weight transfer. The ability of our results to reproduce the complex experimental picture strongly supports the physical picture where the broken-symmetry phases observed in iron-based superconductors and other Hund's correlated metals can only be understood in terms of instabilities of the Hund's metal and a successful theory of these phenomena should include the dynamical correlation effects characteristic of the Hund's metal.

\section*{ACKNOWLEDGEMENTS}
We are grateful to H.~Pfau for helpful discussions. L.~F. acknowledges financial support from the European Union Horizon 2020 research and innovation programme under the Marie Sklodowska-Curie grant SuperCoop (Grant No 838526). A.~V. acknowledges financial support from the Austrian Science Fund (FWF) through through project P 31631. M.C. acknowledges  financial  support from  MIUR through the PRIN 2017 (Prot. 20172H2SC4 005) programs and Horizon 2020 through the ERC project FIRSTORM (Grant Agreement 692670).

%\bibliography{Iron_SC}

\section{Supplementary Material}
\beginsupplement

\subsection{Model}

We consider a three orbital tight-binding model adapted from \cite{Daghofer_PRB2010} already used in \cite{Fanfarillo_PRL2020}. It reproduces qualitatively the Fermi surfaces (FS) typical of the iron-based superconductors (FeSC) family: a two hole-like pockets composed by $yz$-$xz$ orbitals around the $\G$ point and two elliptical electron-like pockets formed by $xy$ and $yz/xz$ orbitals centered at the $X/Y$ point of the 1Fe-BZ. The Hamiltonian reads 
\be
\label{eq:HTB}
 H_{K} = \sum_{{\bf k}\sigma\mu\nu} T^{\mu\nu}({\bf k})
          c^{\dagger}_{{\bf k}\mu\sigma} c^{\phantom{\dagger}}_{{\bf k}\nu\sigma}
\ee
$\mu,\nu$ are orbital indices for the $1=yz$, $2=xz$, $3=xy$ orbitals, $c^{\dagger}_{{\bf k}\mu\sigma}$ ($c^{\phantom{\dagger}}_{{\bf k}\nu\sigma}$) is the fermionic operators that creates (annihilates) an electron 
in orbital $\mu$, with momentum ${\bf k}$ and spin $\sigma$. The intraorbital dispersion are
\bea
 T^{11}&=& 2 t_{2} \cos(k_xa) + 2 t_{1} \cos(k_ya) + \nn \\
 &+& 4 t_3 \cos(k_xa)\cos(k_ya)-\mu,
\lb{eq:Tii}
\eea
\bea
T^{33}&=& 2 t_5 (\cos(x_xa)+\cos(k_ya)) + \nn \\
&+& 4 t_6\cos(k_xa)\cos(k_ya)-\mu +\Delta_{xy},
\lb{eq:T33}
\eea
and $T^{22} = T^{11}$ with $k_x \leftrightarrow k_y$. The interorbital dispersion are given by
\bea
T^{12}=T^{21}= 4 t_4\sin(k_x)\sin(k_y),
\lb{eq:T12}
\eea
\bea
T^{13} = (T^{31})^* &=& 2\imath t_7 \sin(k_xa) \nn \\ 
&+& 4\imath t_8 \sin(k_xa)\cos(k_ya), 
\lb{eq:T13}
\eea
\bea
T^{23}=(T^{32})^*= 2\imath t_7 \sin(k_ya) + 4\imath t_8 \sin(k_ya)\cos(k_xa).
\lb{eq:T23}
\eea
The hopping parameters (in units of eV) are: $t_1=0.02$, $t_2=0.06$, $t_3=0.03$, $t_4=-0.01$, $t_5=0.1$, $t_6=0.15$, $t_7=-0.1$, $t_8=-t_7/2$, $\Delta_{xy}=0.2$. 

We further introduce a nematic perturbation to the above Hamiltonian. We consider an on-site ferro-orbital (OFO) splitting which lifts the degeneracy of the $xz/yz$ orbitals
\begin{equation} \label{eq:HOFO}
 \delta H^{OFO} = - \sum_{\bf k} [n_{xz}({\bf k}) - n_{yz}({\bf k})] \delta \epsilon,
\end{equation}
where $n_{\mu}({\bf k})$ is the number operator in momentum space, and $\delta\epsilon$ the magnitude of the nematic perturbation. The sign of the perturbation was chosen to qualitatively reproduce the hierarchy of the splitting and of the orbital coherence observed in recent photoemission experiments~\cite{Pfau_PRB2021_122,Pfau_PRB2021_FeSe}. 

\begin{figure}[t]
\centering
\includegraphics[width=1.0\linewidth, angle=0]{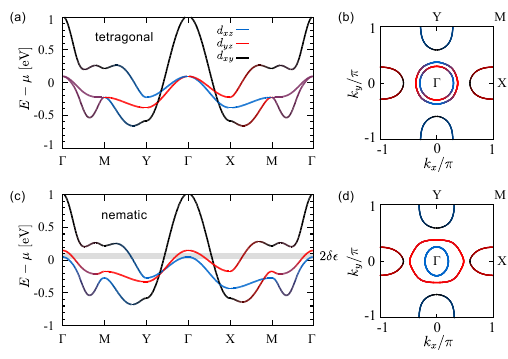}
\caption{Non interacting model. Bandstructure and Fermi surface of the tight-binding model from Eq.~(\ref{eq:HTB}-\ref{eq:T23}) without (a,b) and with (c,d) the bare nematic OFO perturbation  Eq.~(\ref{eq:HOFO}). The colors describe the weight of each orbital within the bands. The grey area in (c) highlights the bare nematic splitting $2\delta\epsilon$, for $\delta\epsilon=50$~meV.}
\label{fig:bands_fsurf}
\vspace{-0.3cm}
\end{figure}

Fig.~\ref{fig:bands_fsurf} shows the electronic bandstructure of the tight-binging model along a high-symmetry path in the 1Fe-BZ that highlights the differences between the tetragonal and nematic phases. The perturbation lifts the $xz/yz$ band degeneracy at the $\Gamma$ point, and it induces a momentum-dependent deformation of the FS.

Local electronic interactions are included in the model considering the multiorbital Kanamori Hamiltonian 
\bea
 H_{int}&=& \frac{U}{2}\sum_{i \mu \sigma} n_{i \mu \sigma} n_{i \mu \bar{\sigma}} 
 + \frac{U^\prime}{2}\sum_{\substack{i \mu \neq \nu \\ \sigma  \tilde{\sigma}}} n_{i \mu \sigma} n_{i \nu 
 \tilde{\sigma}} +\nonumber \\
 && + \frac{J_H}{2}\sum_{\substack{i \mu \neq \nu \\ \sigma  \tilde{\sigma}}} c^{\dagger}_{i \mu \sigma}  c^{\dagger}_{i \nu \tilde{\sigma}} c^{\phantom{\dagger}}_{i \mu \tilde{\sigma}} c^{\phantom{\dagger}}_{i \nu \sigma} +\nonumber \\
 && + \frac{J_H}{2}\sum_{i \mu \neq \nu \sigma} c^{\dagger}_{i \mu \sigma}  c^{\dagger}_{i \nu \bar{\sigma}} c^{\phantom{\dagger}}_{i \nu \bar{\sigma}} c^{\phantom{\dagger}}_{i \nu \sigma} 
\eea
where $n_{i \mu \sigma}=c^{\dagger}_{i \mu \sigma} c^{\phantom{\dagger}}_{i \mu \sigma}$ is the density operator. $U$ and $U'$ are the intraorbital and interorbital Hubbard interactions, $J_H$ is the Hund's coupling. We assume the system to be rotationally invariant, and thus $U'=U-2J_H$ \cite{Castellani_PRB1978}.

We do not include spin-orbit coupling in our minimal model to reduce the number of model parameters and make the interpretation of our theoretical results in term of correlations easier. The spin-orbit coupling is an important ingredient in FeSC modeling, especially if one is interested in the analysis of the electronic structure. However, neither theory nor experiments found a crucial role of the spin-orbit coupling in affecting qualitatively the nematic order in Fe-SC. In particular, recent ARPES experiments on 122~\cite{Pfau_PRB2021_122} and FeSe~\cite{Pfau_PRB2021_FeSe} give the same qualitative results despite the large difference in the spin-orbit coupling between the two compounds.

\subsection{Dynamical Mean-Field Theory}

The effect of the interactions is analyzed within the Dynamical Mean-Field Theory (DMFT) approximation~\cite{DMFT_review}. 
For fixed values of $U$ and $J_H$, we tune the chemical potential to set the filling to $n=4$ electrons, and we compute the local self-energy $\Sigma_{\mu\mu}(i \omega_n)$, where $\mu$ in the orbital index and $\omega_n$ is the $n$-th fermionic Matsubara frequency. 
Within this scheme, the ${\bf k}$-resolved spectral function is obtained from the retarded Green's function as 
\begin{equation}
 A_{\mu}({\bf k},\omega) = -\frac{1}{\pi} \Im G_{\mu\mu}({\bf k},\omega), 
\end{equation}
where
\begin{equation}
 G_{\mu\nu}({\bf k},\omega) = \Big[ (\omega+\imath\eta) \delta_{\mu\nu} - H_{\mu\nu}({\bf k}) - \Sigma_{\mu\nu}(\omega) \Big]^{-1}.
 \end{equation}
Note that since the local Hamiltonian of the model in Eqs.~(\ref{eq:HTB}-\ref{eq:T23}) does not have interorbital terms (local hybridizations), the interorbital elements of the self-energy vanish, i.e., $\Sigma_{\mu\neq\nu}(\imath\omega_n)=0$ and the self-energy is given by $\Sigma(\omega)=\text{diag}(\Sigma_{11},\Sigma_{22},\Sigma_{33})$. 

In order to compare our results to photoemission spectra~\cite{Pfau_PRB2021_122,Pfau_PRB2021_FeSe}, we perform a partial integration of the spectral function along specific paths in the 1Fe-BZ. In the text we indicate the integrated spectral functions as $A_{\mu}^{\Gamma-X/Y}$ for $yz$ and $xz$, respectively. Specifically, we compute, for the $xz$ orbital
\begin{equation}
 A_{xz}(\omega) = \int_{\Gamma}^{Y} d{\bf k} \ A_{xz}({\bf k},\omega),
\end{equation}
and analogously for the $yz$ orbital 
\begin{equation}
 A_{yz}(\omega) = \int_{\Gamma}^{X} d{\bf k} \ A_{yz}({\bf k},\omega).
\end{equation}

For the auxiliary impurity problem of DMFT, we use an exact diagonalization solver at zero temperature.~\cite{Weber_PRB2012,Amaricci_CPC2022} 
We employ $n_b=3$ bath levels for each impurity orbital, i.e., $n_s = 3 \times (1 + n_b) = 12$, which ensures a good balance between reasonable computational costs and numerical accuracy~\cite{Liebsch_2011}, as already done in previous works~\cite{Fanfarillo_PRL2020}.

\bibliography{Iron_SC}

\end{document}